\documentclass{article}
\usepackage{arxiv}

\usepackage[utf8]{inputenc} % allow utf-8 input
\usepackage[T1]{fontenc}    % use 8-bit T1 fonts
\usepackage{hyperref}       % hyperlinks
\usepackage{url}            % simple URL typesetting
\usepackage{booktabs}       % professional-quality tables
\usepackage{amsfonts}       % blackboard math symbols
\usepackage{nicefrac}       % compact symbols for 1/2, etc.
\usepackage{microtype}      % microtypography
\usepackage{lipsum}		% Can be removed after putting your text content
\usepackage{graphicx} 
\usepackage[acronym]{glossaries}
\usepackage[locale=US]{siunitx}
\usepackage{multicol}
\usepackage{multirow}

\title{Evaluating Different Machine Learning Techniques as Surrogate for Low Voltage Grids}
\date{June 19, 2020}

\author{%
	\href{https://orcid.org/0000-0002-2018-1078}{\includegraphics[scale=0.06]{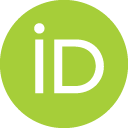}\hspace{1mm}Stephan~Balduin}\\
	OFFIS -- Institute for Information Technology\\
	Escherweg 2, 26121 Oldenburg \\
	\texttt{stephan.balduin@offis.de} \\
\And
	\href{https://orcid.org/0000-0002-8716-0554}{\includegraphics[scale=0.06]{orcid.png}\hspace{1mm}Tom~Westermann} \\
	\texttt{tom.westermann@outlook.de} \\
\And
	\href{https://orcid.org/0000-0003-3796-8931}{\includegraphics[scale=0.06]{orcid.png}\hspace{1mm}Erika~Puiutta} \\
	OFFIS -- Institute for Information Technology\\
	Escherweg 2, 26121 Oldenburg \\
	\texttt{erika.puiutta@offis.de} \\
}

\renewcommand{\headeright}{Preprint. Under Review}
\renewcommand{\undertitle}{A preprint}
\newacronym{vmpu}{vm\_pu}{voltage magnitude per unit}

\hypersetup{
pdftitle={Evaluating Different Machine Learning Techniques as Surrogate for Low Voltage Grids},
pdfsubject={Surrogate models},
pdfauthor={Stephan~Balduin, Tom~Westermann, Erika~Puiutta},
pdfkeywords={machine learning, artificial neural network, surrogate model, power grid, power flow},
}

\begin{document}
\maketitle

\begin{abstract}
The transition of the power grid requires new technologies and methodologies, which can only be developed and tested in simulations.
Especially larger simulation setups with many levels of detail can become quite slow. 
Therefore, the number of possible simulation evaluations decreases.
One solution to overcome this issue is to use surrogate models, i.\,e., data-driven approximations of (sub)systems.
In a recent work, a surrogate model for a low voltage grid was built using artificial neural networks, which achieved satisfying results.
However, there were still open questions regarding the assumptions and simplifications made.
In this paper, we present the results of our ongoing research, which answer some of these question. 
We compare different machine learning algorithms as surrogate models and exchange the grid topology and size.
In a set of experiments, we show that algorithms based on linear regression and artificial neural networks yield the best results independent of the grid topology. 
Furthermore, adding volatile energy generation and a variable phase angle does not decrease the quality of the surrogate models.
\end{abstract}


\section{Introduction}
\label{sec:introduction}

The ongoing transformation of the power system requires the involvement of new technologies and methodologies to meet the requirements that arise during this process.
Since the power grid is a safety-critical infrastructure, the possibilities to test new technologies are very restricted and, since it is also an expensive infrastructure, the installation of a large-scale test grid is not feasible.
For this reason, simulation and hardware-in-the-loop are used for the development and testing of such new technologies \cite{steinbrink2017simulation}.
The so-called smart grid comprises not only the power grid, but also other domains, such as the Information and Communication Technology (ICT) domain and the gas and heat energy systems.
Usually, for each domain, experts are developing their own simulation environment.
One smart solution to couple these different domains incorporates the use of co-simulation, which only requires to implement interfaces for each domain \cite{steinbrink2016nonintrusive}.
In such a setup, synchronization and data exchange between different environments is handled by a co-simulation framework (e.\,g., the co-simulation framework mosaik\footnote{\url{https://mosaik.offis.de}, \footnotesize{retrieved on 10 Jun. 2020}}).
% par

Even with co-simulation, building such a large, cross-domain simulation environment is still a complex task.
In addition, depending on the size and complexity of the underlying scenario, the simulation of the overall system can become very slow \cite{Bla2015}.
To overcome this, surrogate models can be used to reduce the simulation time of some of the components in the environment.
Usually, a surrogate model is a data-driven approximation of a certain function or system, which can be evaluated faster than the original function or system, but which is less accurate \cite{Simpson2001}.
% par

In a recent work, a surrogate model for a subsystem of such a multi-domain co-simulation environment was built \cite{balduin2019towards}. Starting with the Conseil International des Grands Réseaux Électriques (CIGRE low voltage (LV) benchmark power grid model \cite{papathanassiou2005benchmark}, the authors added several load models that were connected to certain buses in the grid.
A surrogate model was built to replace the load models and the power grid model.
The authors evaluated the surrogate model in a case study with several experiments.
Their goal was to enable the creation of larger setups of medium voltage (MV) grid simulations using this kind of surrogate model as a replacement for some of the LV grids.
% par

Although we think this is a promising approach, the authors concluded their work with several simplifying assumptions and open questions, from which we recap the most important ones.
First, the simulation setup consisted only of load models, but in a smart grid scenario, there would also be distributed energy resources (DER) like photovoltaic (PV) or combined heat and power (CHP).
Secondly, the authors described the architecture of their deep neural network (DNN) surrogate model as simple, which we will discuss later in this paper.
Furthermore, other machine learning (ML) techniques could be better suited to this particular problem.
Finally, although the proposed methodology worked for the CIGRE LV grid, it would be interesting to know if this methodology works for other grid models, as well.
% par

In this paper, we aim to address these issues. 
In the next section, we give an overview of \nameref{sec:relatedwork} done in this field.
This is followed by a description of the \nameref{sec:simsetup} created in the previous work that served as the basis for our study.
In addition, we discuss the modifications we made to this simulation setup.
In the subsequent section, we present the \nameref{sec:casestudy} of this work, comprising hypotheses, experiments, and the presentation of the results.
We conclude our work and discuss the results in the section \nameref{sec:conclusion}.
Finally, we discuss open questions and future work in the section \nameref{sec:outlook}.

\section{Related Work}
\label{sec:relatedwork}

Surrogate models -- in the literature also referred to as metamodels, response surfaces, or emulators -- are a common technique from the field of statistical design of experiments \cite{Myers2016}.
They are used to describe the behavior of a system that, for various reasons, is not suited to be built knowledge-based.
Surrogate models can be found in many domains, but we will focus on the energy domain.
They are used in a broad range of use cases: 
starting from the calculation and optimization of energy savings \cite{beisheim2019energy, vazquez2019deep, nagpal2019methodology} and the replacement of specific simulation models \cite{dimitrov2019surrogate, balduin2019tool} over surrogate models for (micro)grids \cite{grundel2019surrogate, baumann2019surrogate, balduin2018surrogate} to the use in uncertainty and reliability assessment \cite{slot2020surrogate, blank2014correlations, steinbrink2016nonintrusive}.
This list is far from complete and there are also other approaches such as in Gerster \cite{gerster2018agent} who use surrogate models to build a decoder function abstracting from technical system specifications. 
% par

When used for grid emulation, surrogate models are basically used to replace a power flow (PF) analysis. 
PF analysis is performed in different use cases; market analysis and short-term operational planning are only two of them.
The PF analysis part used in this paper assumes that the power grid is in a steady-state and has no special use case in mind despite the PF analysis itself.
The most common traditional methods for PF are Newton-Raphson (NR), Gauss-Seidel (GS), and derivatives, which calculate bus voltages, currents, etc. numerically.
These methods have some important drawbacks. 
Most of them require to perform matrix inversions, which are solved iteratively.
Bad initial guessing of unknown values in this process can lead to divergence and, subsequently, a repetition is required.
Therefore, other solutions to this problem are actively researched.
% par

Grisales-Nore{\~n}a et al. \cite{grisales2020application} compared six different methods from the literature with their approach of backward/forward sweep iteration for the PF calculation in direct current (DC) grids with radial structure.
The authors benchmarked their approach on four different grid sizes and achieved a performance increase, especially on the larger grids.
Another approach from Montoya et al. \cite{montoya2020power} aimed to overcome the issue of the costly matrix inversions that are normally required to solve the PF.  
Montoya et al. proposed a classical gradient conjugate method to solve linear algebraic equations in DC grids.
Kontis et al. \cite{kontis2019power} investigated the PF in islanded DC microgrids that are operated under the droop control scheme.
A steady-state analysis is not easy to employ for islanded grids since there is no slack bus, but the proposed method showed promising results regarding accuracy and robustness.
Yuan et al. \cite{yuan2019fast} proposed a linear redefinition of the nonlinear power flow equations for DC grids. 
Although not applicable to alternating current (AC) grids, the authors reported a calculation speed twenty times faster than using NR on grids with more than a hundred nodes and an extremely small error.
% par

But there is also a broad variety of approaches to solve the PF and related problems with data-driven algorithms.
Nilsson et al. \cite{nilsson2018machine} built a ML-based simplified grid model to perform a PF analysis. The authors concluded that their model was good enough to be used in several applications such as security-constraint dispatch and intra-hour simulations.
The work in \cite{veerasamy2020novel} created a modified Hopfield artificial neural network (ANN) to solve PF equations. 
The results were evaluated on different standard IEEE power systems.
Another example is given by \cite{donon2019neural} who proposed an ANN based on a Graph Neural Solver.
Frank et al. \cite{frank2012optimal} surveyed about data-driven approaches to the closely related optimal power flow (OPF) problem.
In \cite{xiang2020probabilistic}, the authors used a DNN to overcome the high computational burden of OPF.
Syai'in and Soeprijanto \cite{syai2010neural} proposed their method called ANN OPF, compared it to improved particle swarm optimization, and achieved a faster response time.
Gupta et al. \cite{sudha2015neural} used a neural network to predict cascading failures in advance.
% par 

One question when applying ML to power grid models is related to the gathering of training data.
In \cite{balduin2019towards}, the authors pointed out the challenges of the sampling process for power grids.
They tried to replace their set of load profiles using an empirical sampling strategy with kernel density estimation, but this method did not provide satisfactory results.
Therefore, the authors did not fully replace their initial data but extended them with sampled data. 
Danner and de Meer \cite{danner2019state} discussed several methods to perform state estimation in the LV distribution grid.
To solve the sampling problem, the authors compared Monte-Carlo simulation with one-at-a-time sensitivity analysis and realistic load profiles. 
Danner and de Meer used the sampled data to build a dependency graph, which is used as input for different ML techniques.
Although they did not present final results, their proposed methodology sounds interesting and worthy to investigate further.
% par

A quite different approach is given by Zhao et al. \cite{zhao2019power}. 
They compared techniques from the fields of model reduction and machine learning, which are quite similar but with subtle differences, to build parametric surrogate models.
The two case studies presented by Zhao et al. considered the prediction of pressure fields around an airfoil and the prediction of strain field over a damaged composite panel. 
The authors showed that such a parametric surrogate model can reduce the dimensionality of the model and  allows to embed physical constraints, which were required for their case studies. 
Though their work is not located in the energy domain, the approach could also be useful to mitigate the sampling problem for power grids.
As a further benefit, Zhao et al. argued that their model could also improve the interpretability of the trained models.
Interpretability and explainability are a current issue when it comes to ML models that are used in critical decision making, especially for deep learning (DL) and reinforcement learning (RL).
For a deeper insight in the field of explainability of AI and RL we refer to \cite{puiutta2020xrl}.
% par

Another possible approach would be the use of RL to explore the grid and to aid in the sampling process.
The concept of adversarial resilience learning (ARL) \cite{fischer2019arl} utilizes two classes of RL agents competing in a shared environment.
In the simplest case, one agent tries to break the environment, and the other one aims to keep the environment in a healthy state \cite{veith2020adversarial}.
The ARL methodology could be used to explore the sampling space, localize limitations of the model, and identify critical spots to derive sampling rules. 

\section{Simulation Setup}
\label{sec:simsetup}

The simulation setup in \cite{balduin2019towards} used the pandapower \cite{pandapower2018} implementation of the CIGRE LV benchmark grid \cite{papathanassiou2005benchmark}.
This grid consisted of three subgrids: a residential subgrid, an industrial subgrid, and a commercial subgrid.
A time series was assigned to each of the load models. 
In each step, these models forwarded the corresponding value from the time series to the grid.
For the residential area, a data set from the research project Smart Nord \cite{Mar2015} was used that consisted of synthesized time series of households.
In total, these time series approximated the default load profile that is commonly used in Germany to approximate household consumer load.
For the industrial and commercial areas, a reference data set from openei\footnote{\url{https://openei.org/datasets/files/961/pub}, \footnotesize{retrieved on 10 Jun. 2020}} was used, which consisted of time series of several commercial consumers.
% par

\begin{figure}
	\centering
	\includegraphics[width=0.9\textwidth]{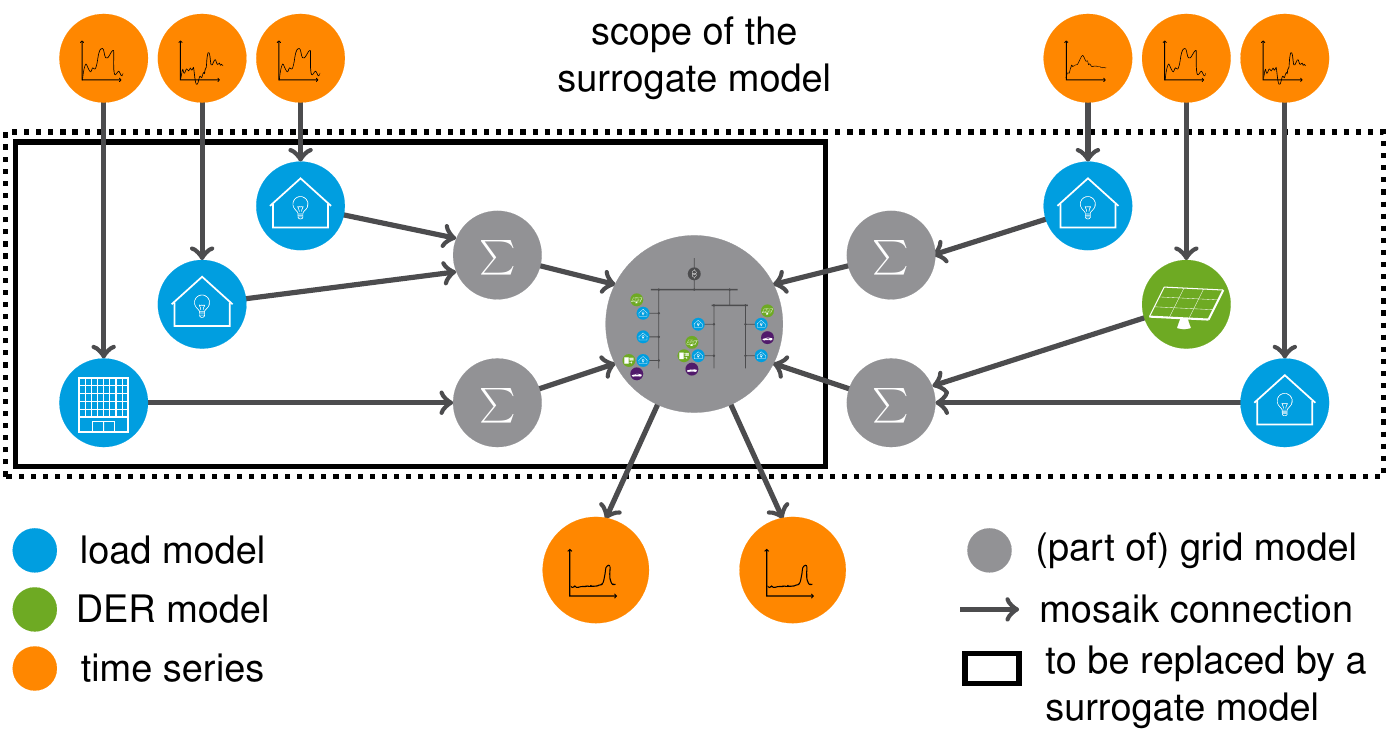}
	\caption{Simulation setup used in the referenced work is represented by the left part. Data from time series (orange circles) were transmitted to the load models (blue circles), from the load models to one of the load aggregators (grey circle with the $\sum$ symbol), and from there to the grid model (large grey circle). The grid model performed PF calculations and the results were gathered, again, as a time series. In the referenced work, the models inside the bold black box were replaced by a surrogate model. For this work, we extended the scenario for the right part and built surrogate models for the models inside the dashed black box.}
	\label{fig:architecture}
\end{figure}

The load models and the pandapower grid model were coupled using the co-simulation framework mosaik.
While in the industrial and commercial subgrids to each load  of the grid one load model was connected to, in the residential subgrid several load models could be connected to the same load node.
This assignment was considered as domain knowledge about the grid as well as other topological information such as which load node is connected to which bus.
The surrogate model has not received this information, but only time series as input, i.\,e., in this regard, the model was built without domain knowledge.
\autoref{fig:architecture} illustrates this relationship and shows what is and what is not part of the surrogate model.
This decision -- to distinguish between time series and a value-forwarding load model -- was made with the goal to incorporate PVs and CHPs as part of this system to-be-replaced as well.
Using unassigned load data and, in the future, the weather forecast for DER, the surrogate model should predict the \gls*{vmpu} of the grid's buses.
In a larger setup, this would add the value of a better estimation and more flexibility of this LV grid regarding distributed energy generation in comparison to a setup where this grid is replaced by a single time series.

The surrogate model was built using a DNN, which consists of several fully connected (dense) hidden layers. 
A random-search cross-validation approach was used to evaluate a small set of hyperparameters.
Among others, this included the number of hidden layers.
However, the number of neurons in each hidden layer was determined depending on the sizes of input- and output-layer as well as the number of hidden layers, based on a rule-of-thumb \cite{heaton2008introduction}.
The resulting model was tested in different experiments evaluating the accuracy, a speed-up factor (SUF), and the capability to correctly predict voltage violations at the grid's buses.
Since the entire architecture was created generically with few, domain-independent parameters, the authors concluded that the model architecture was simple and should be further improved.
% par

For this work, we aimed to put some domain knowledge into the DNN and compared the new model to the reference model from the previous work.
Furthermore, we evaluated additional ML models to find the most suitable model for this simulation setup.
Therefore, we included ML models with different characteristics ranging from single target predictor ensembles to multi-target models and recurrent neural networks (RNN), resulting in a total of six different surrogate models.
We discuss these models further in section \nameref{sec:modelselection}.
All models were compared in experiments similar to the experiments in the reference work.
% par

The second goal of this work was to evaluate this methodology on a different simulation setup.
The simbench project \cite{spalthoff2019simbench} provides data sets for power grid benchmarks, including certain grid topologies.
The \emph{LV-rural3} data set describes a LV grid including time series of household loads and PV power generation. 
The simbench grids are available for pandapower which enabled us to extend the architecture without much additional work.
Using this simbench grid, we aimed to address the open questions of the previous work.
% par

\begin{figure}
	\centering
	\includegraphics[width=0.9\textwidth]{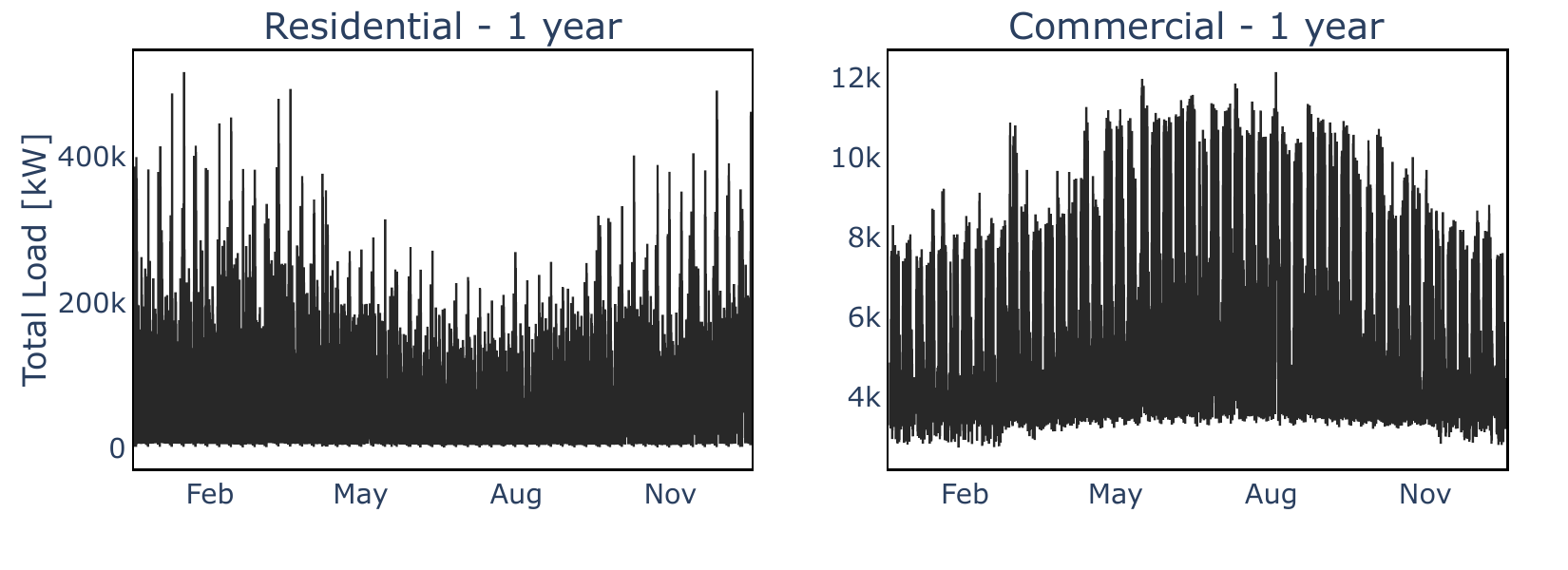}
	\caption{Aggregation of the loads used in the CIGRE LV grid. The industrial load is included in the right graph.}
	\label{fig:datacigrelv}
\end{figure}

\begin{figure}
	\centering
	\includegraphics[width=0.9\textwidth]{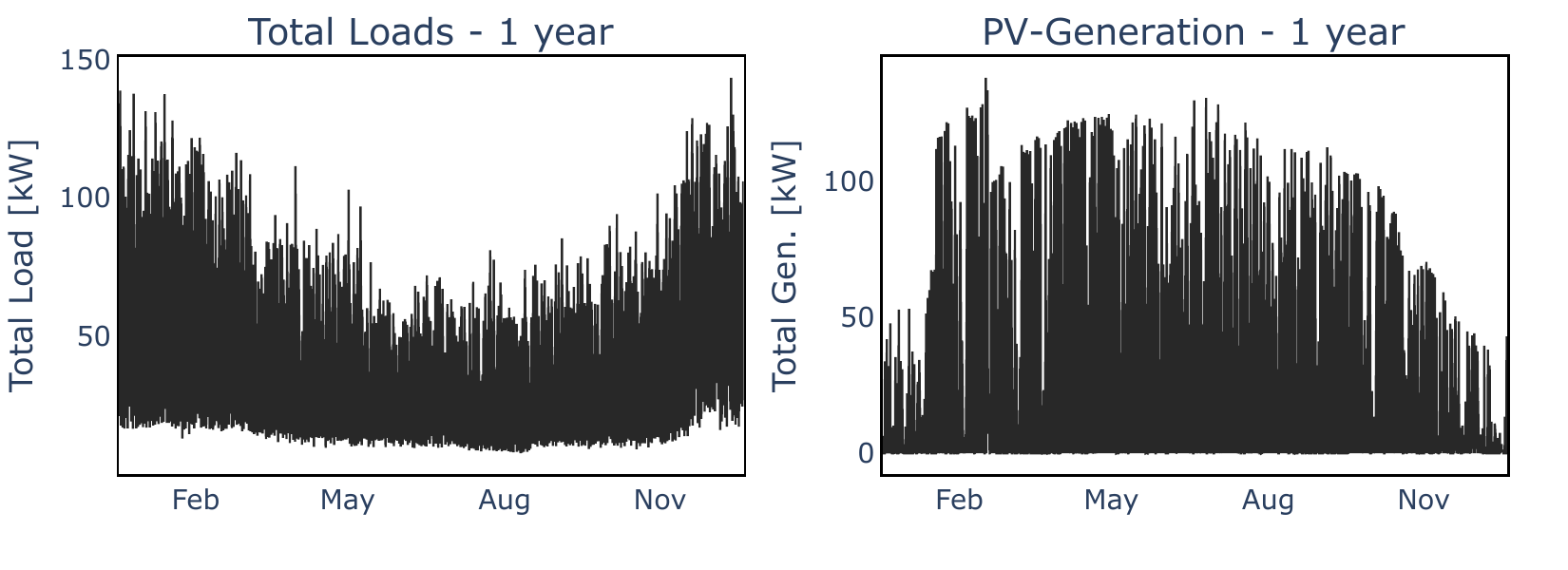}
	\caption{Aggregation of the load and generation used in the LV-rural3 grid.}
	\label{fig:datalvrural}
\end{figure}

Beforehand, we analyzed and compared both data sets.
Each time series contains 15-minute averages over one year resulting in a total of 35,400 entries. 
\autoref{fig:datacigrelv} shows the aggregation of the residential time series on the left and the commercial time series on the right. 
The residential loads tend to be higher in the cold months, lower in summer, and have noticeable fluctuations. 
A more regular behavior can be seen at the commercial and industrial loads, which, unlike the residential loads, tend to be higher in summer.
The data of the simbench project are shown in \autoref{fig:datalvrural}.
On the left are, again, the residential loads.
These show a behavior similar to the Smart Nord loads, with the tendency to be higher in winter.
Instead of commercial loads, the simbench project provides time series for PV generation, which are shown on the right.
The PV generation is higher in summer, remarkably low in December and January, and heavily depending on the weather conditions.

\begin{figure}[t!]
	\centering
	\includegraphics[width=0.9\textwidth]{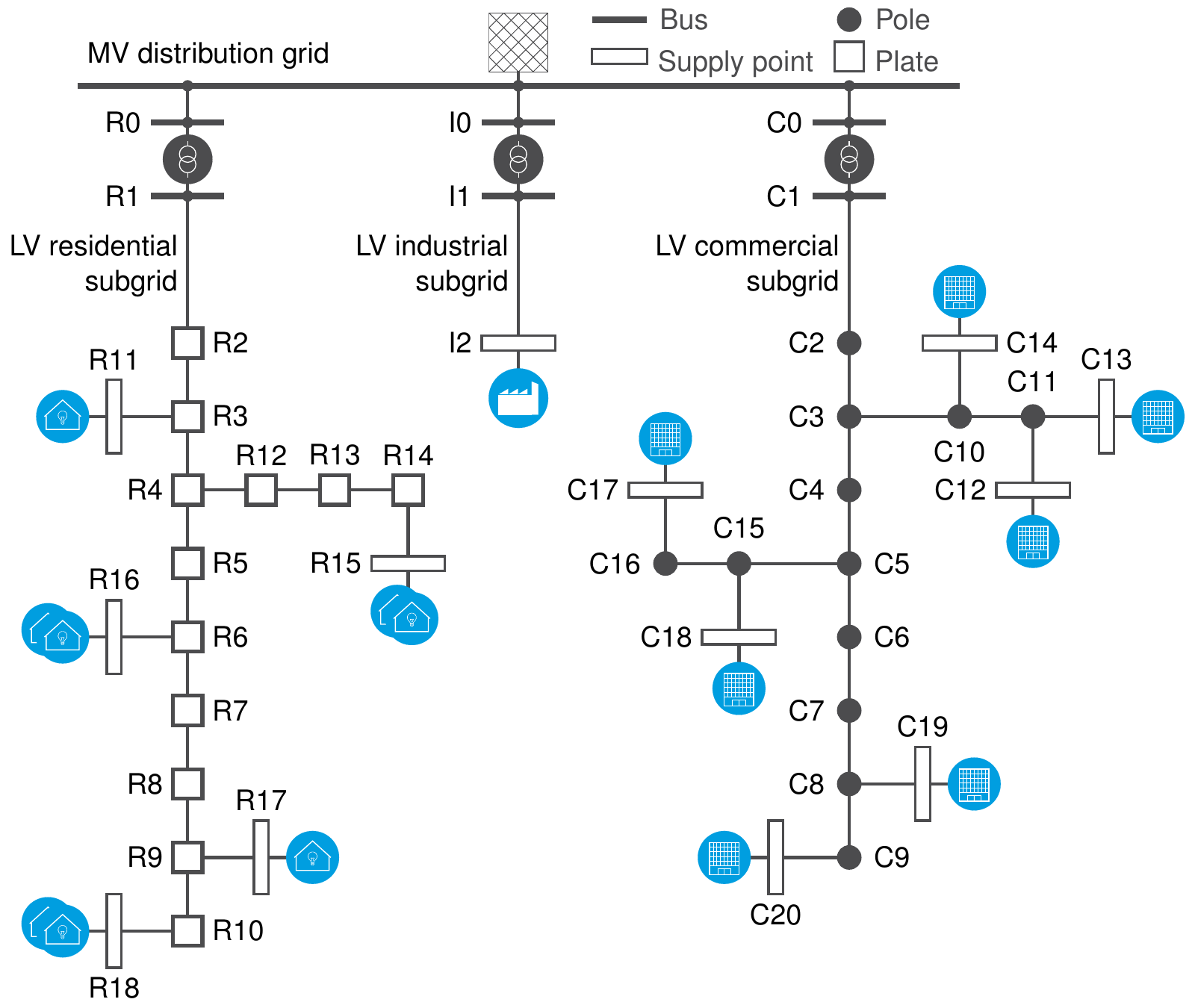}
	\caption{Topology of the CIGRE LV benchmark grid, from \cite{balduin2019towards}. The blue boxes represent load nodes, with different load types indicated by the symbol inside the box. Lines represent lines of the grid, except for the bold lines, which represent bus bars.}
	\label{fig:cigretop}
\end{figure}

\begin{figure}
	\centering
	\includegraphics[width=0.9\textwidth]{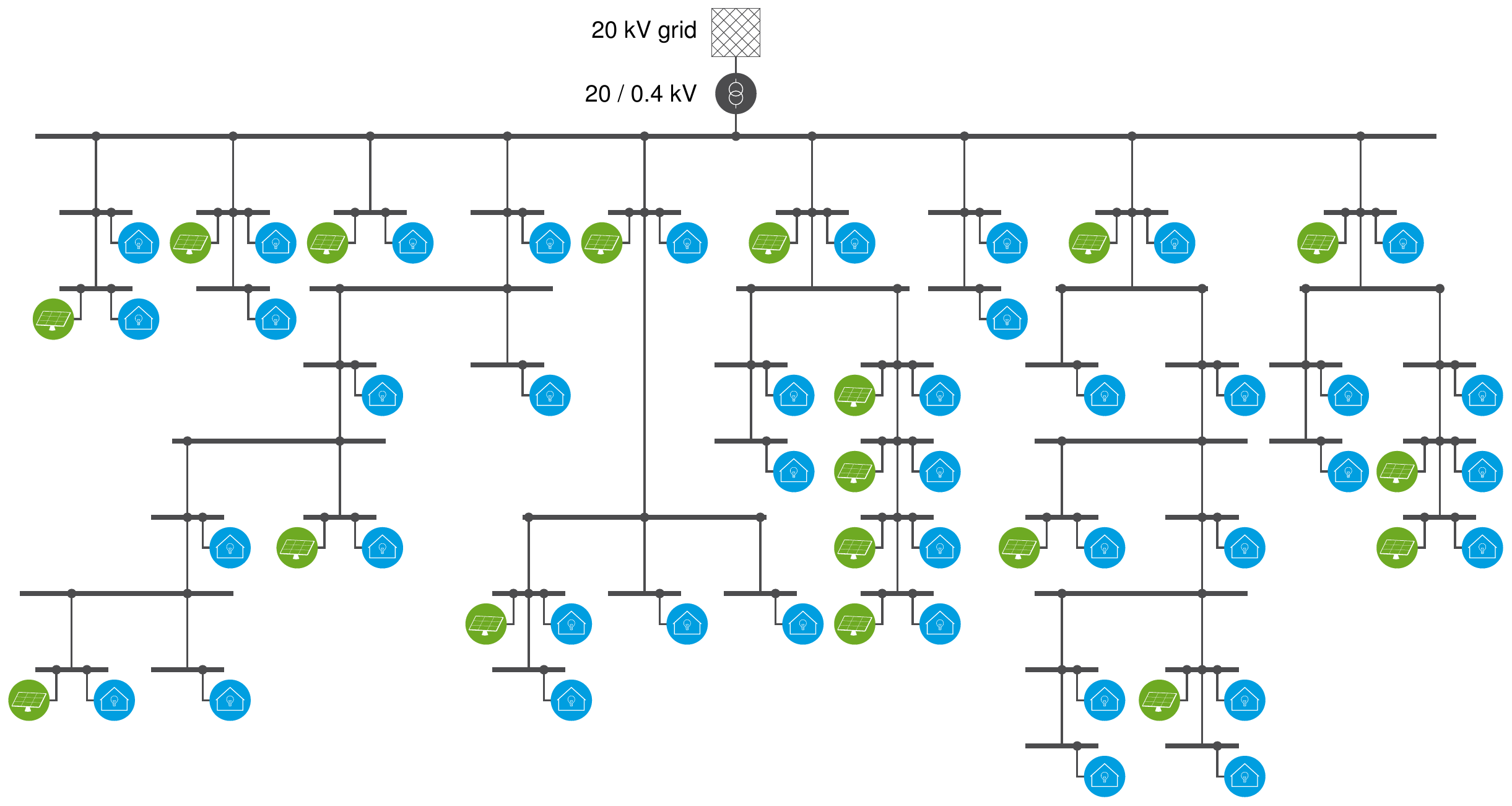}
	\caption{Topology of the simbench LV-rural3 grid, inspired by \cite{spalthoff2019simbenchdoku}. The blue boxes represent load nodes and the green boxes represent distributed energy generation. }
	\label{fig:simbenchtop}
\end{figure}

Finally, we compared the different topologies, which are illustrated in \autoref{fig:cigretop} and \autoref{fig:simbenchtop}. 
The CIGRE LV grid includes 44 buses and 15 loads.
These are distributed over the three subgrids mentioned before.
The LV-rural3 has considerably more components: 128 buses, 118 loads, and 17 PV plants.
In contrast to the CIGRE grid, LV-rural3 includes time series for active and reactive power, providing a varying phase angle.

\section{Case Study}
\label{sec:casestudy}

Based on this setup, we conducted a case study to validate our extensions against the open questions of the previous work.
We will discuss these questions in the first part of this section and then hypothesize to define the context of our experiments.
The second part of this section consists of a brief overview of the ML models, their parameter tuning strategies, and the reasons for their selection.
Afterwards, we describe the data generation and the experimental setup in detail.
Finally, this section concludes with a description of  the experiments' results.

\subsection{Hypothesis}
\label{sec:hypothesis}

The work, this study is based on, aimed to provide a benchmark model and evaluation environment for further experiments as a proof-of-concept. 
This explains why (a) the architecture of the ANN was rather simple and (b) there was no reasoned selection of ANNs as a surrogate model in general.
We aimed to provide a well-founded basis for these crucial points, selected a variety of models, and improved the hyperparameter tuning (see \nameref{sec:modelselection}).
Hereafter, any mention of a part of the study (e.\,g., hypothesis, or surrogate model) is referring to the corresponding part in the reference \cite{balduin2019towards}.
% par

To compare these new models to the reference model, we defined three hypotheses similar to the reference hypothesis. 
The first one deals with the prediction quality of the models. 
To be predicted were the \gls*{vmpu} of the buses, which were normalized values that should lie in the interval [0.9, 1.1].
In the reference work, a quality threshold of \SI{10}{\percent} was used but the authors concluded that the model was not accurate enough.
Therefore, we will set the threshold to \SI{0.1}{\percent}.
The reasoning behind this decision as well as the used metric for the error are discussed in the section \nameref{sec:exp1descr}.

\paragraph*{Hypothesis 1: Surrogate Model Accuracy}
The behaviour of the voltage magnitudes of a low-voltage power grid simulation model can be adequately captured by a surrogate model.

\begin{itemize}
	\item[$H_0$] error of the models $>$ \SI{0.1}{\percent}
	\item[$H_a$] error of the models $\leq$ \SI{0.1}{\percent}
\end{itemize}

The main purpose of using surrogate models is to reduce the complexity of the original system.
Therefore, the second hypothesis dealt with the question of whether the surrogate models actually reduced the computation time required for the simulation. 
Since we did not know in advance if the reference model was exemplarily good or bad regarding the computation time and compared to other models, we compared all surrogate models with the original simulation model.
For this experiment, we will defined an arbitrary time frame that was the same for all models.
Each model repeatedly had to calculate the values for this time frame.
The averaged results were compared to each other.

\paragraph*{Hypothesis 2: Calculation Speed}
The surrogate models‘ calculation time $t_{sur}$ on a defined time frame differs significantly from the simulation models’ calculation time $t_{sim}$. 

\begin{itemize}
	\item [$H_0$] $t_{sur}$ = $t_{sim}$ ($\alpha \geq 0.05$)
	\item [$H_a$] $t_{sur}$ $\neq$ $t_{sim}$ ($\alpha < 0.05$)
\end{itemize}

In the third hypothesis, we verified whether our findings can be transferred to other grid topologies as well.
Additionally, several simplifications were made for the reference environment.
Only load models were used and, though with different characteristics, a constant phase angle was used.
We selected the simbench LV-rural3 data set since it provides PV power generation and a varying phase angle as well as a totally different grid topology.
We discuss the difference in more detail in the section \nameref{sec:datageneration}.
With LV-rural3, some of the simplifying assumptions were eliminated and one step toward a generalizability check was done.
However, since both grids are too different to be compared quantitatively, we performed a qualitative comparison. 
We provide more information on this topic in section \nameref{sec:exp2descr}.

\paragraph*{Hypothesis 3: Generalization}
The results concerning the surrogate models' prediction accuracy and their calculation speed can be generalized to other simulation models.

\begin{itemize}
	\item no test criterion, since there are so many variables involved. 
	\item instead, a qualitative comparison between two simulation models will be performed. 
\end{itemize}

\subsection{Model selection}
\label{sec:modelselection}

Our goal was to compare models with different characteristics to provide an overview of which class of models performed best for the given task.
In addition to the reference model, we selected models from four different families of machine learning algorithms.
% par

First, we distinguished between single-target and multi-target models.
Linear regression (LR) belongs to the former class and is a well established and widely known algorithm to build a regression model.
LR requires a fairly low number of calculation steps and, since only one output is calculated, the result is interpretable and explainable.
Another single-target model is the random forest (RF), which is an ensemble method of decision trees.
The RF algorithm is more flexible than regular LR due to it's ability to also model non-linear input-output relationships.
However, large numbers of trees slow down the training process.
In contrast to regular LR, RFs have several hyperparameters.
The three most important ones are candidate variables per split (\emph{mtry}), the \emph{sample size} of each tree's training process, and the minimum number of observations in an endnode (\emph{nodesize}) \cite{probst2019hyperparameters}.
To provide multiple outputs as required by the given task, both models were combined to a regressor ensemble (RE).
Additionally, LR was evaluated in a regressor chain (RC), which is mathematically equivalent, but is potentially more efficient to compute \cite{spyromitros2016multi}.
In total, three single-target models are evaluated -- regressor ensemble linear regression (RE LR), regressor ensemble random forest (RE RF), and regressor chain linear regression (RC LR) -- and their hyperparameters were optimized with random-search cross-validation.
% par

We also selected three multi-target models.
As a distance-based model we chose k-nearest neighbors (k-NN), which is rather simple but often provides good results.
This algorithm has only a few hyperparameters with \emph{k}, the number of neighbors to be considered, being the most prominent one.
These will be optimized with a grid-search cross-validation.
K-NN provides a fast training process and good prediction speed.
The second multi-target model we selected was the long short-term memory (LSTM) network, which is an adaption of ANNs specifically suited to temporal data \cite{hochreiter1991untersuchungen}.
These kinds of networks are able to consider past values in the prediction and have a large number of hyperparameters, thus we decided to use the \emph{hyperopt} hyperparameter optimization.
Finally, we selected the same kind of ANN that was used in the reference work but with a more fine-grained architecture design.
In contrast to the reference model, which consisted solely of fully-connected layers, we added task-specific layers, which forward their activation only to few or even only to one subsequent node.
Furthermore, we inserted dropout layers after each hidden layer.
In addition to number of epochs and the number of hidden layers, which were the only hyperparameters that were optimized in the reference model, we added batch size, different activation functions, a dropout factor, the number of task-specific layers, and the learning rate of the ANNs optimizer as hyperparameters.
With these six models, we were able to make a reasonable assessment as to which models were best suited for the given use case.

\subsection{Experimental setup}
\label{sec:experiment}

The following section deals with the setup of the experiments. 
We start with a description of the generation of training data. 
The section will be followed up by a description of two experiments we conducted to test the three hypotheses that were established earlier.

\subsubsection{Data generation}
\label{sec:datageneration}

To train surrogate models, a sufficiently large training and testing data set was needed. 
Since both the simulation models and matching realistic load time series were available, calculating the \gls*{vmpu} of the bus bars was a straightforward task. 
Instead of implementing an elaborated experimental design that seeks to map the entire input space, the input values were restricted to realistic combinations of input values. 
This marks a deviation from the sampling process used in the reference work.
The input and output values were obtained within the simulation setup described in the section \nameref{sec:simsetup} whereby all loads (active and reactive power) and all \glspl*{vmpu} were logged at every step. 
In this way, the entire span of the time series (365 days) was used to calculate a set of input and output values, which could be used to train the models.
% par

There were a number of differences between the two simulation models that had an effect on the process of data generation. 
Since one of the assumptions for the CIGRE LV was a constant phase angle $cos(\varphi)=0.9$ between the voltage and the current in the grid, the reactive power $Q$ for the input values was calculated from the active power $P$. 
The LV-rural3, on the other hand, had separate time series for active and reactive power consumption, thus it featured a variable phase angle $cos(\varphi)$. 
Additionally, the LV-rural3 included numerous distributed energy generators in the form of PV systems, each of which had its own time series for active and reactive power fed into the grid. 
These time series were also used as input values for the simulation models.

\subsubsection{Description of Experiment 1}
\label{sec:exp1descr}

The goal of the first experiment was to verify whether the surrogate models were able to accurately predict the \glspl*{vmpu} calculated by the simulation model. 
Since we intended to explore the maximum accuracy potential of the machine learning algorithms, a training set of maximum size was used. 
However, since one of the surrogate models was based on a LSTM network, randomly splitting the data into training and testing set eliminates temporal information required by this model. 
The load time series consisted of twelve months of data, of which eleven were used for training and one for evaluation.
Since the data set showed significant differences between different months, the process was repeated twelve times in the form of a 12-fold cross validation. 
Thus, we could evaluate the models over the span of the entire year. 
Due to the lack of a larger data set and the cyclical nature of the data, we used observations from the months following the test month as well.
To quantify the prediction error, the average root mean squared error (RMSE) over all $n$ observations and all $m$ bus \glspl*{vmpu} of the LV grid was calculated.

\begin{equation}\label{eq:rmse}
RMSE = \sqrt{\frac{\mathrm{1} }{\mathrm{n m}}  
	\sum\nolimits_{i=1}^{n}
	\sum\nolimits_{j=1}^{m}
	(y_{ij} - \hat{y}_{ij})^{2} }
\end{equation}

At this, $y_{ij}$  is the actual value and $\hat{y}_{ij}$ the prediction of the \gls*{vmpu} of bus $j$ in observation $i$.
The RMSE was used as a criterion to decide whether the null hypothesis can be rejected. 
The more challenging aspect of this criterion was to determine an adequate threshold value. 
Different applications of the surrogate models could potentially have completely different requirements in terms of accuracy and reliability. 
Furthermore, too high accuracy on the training set may be the consequence of overfitting.
Since the voltage magnitudes were given in the per-unit-system, the observed ranges of voltage magnitudes spans from 0.87 to 1.0 for CIGRE LV and from 0.95 to 1.05 for the LV-rural3.
For this reason, we chose to set the threshold for an adequate prediction error at an RMSE of $10^{-3}$ and skipped further normalization of the error.
This corresponded to approximately one percent of the observed values' range. 
Therefore, we could reject the null hypothesis for a given surrogate model, if the average RMSE over the span of a year is greater than $10^{-3}$ ($\overline{RMSE}_{year} \leq 10^{-3}$).
In order to also be able to judge whether or not the obtained results were robust towards changes in the LV grids' parameters, the experiment was repeated on the LV-rural3 grid. 
We then quantitatively compared and examined the experiments' results to find possible effects caused by the changed parameters.

\subsubsection{Description of Experiment 2}
\label{sec:exp2descr}

The second experiment aimed to provide insight into the surrogate models' calculation speed when used in a co-simulation-setup. 
In order to provide a baseline calculation time, the same experiment was conducted with the simulation models.
Each of the models we built in experiment 1 was part of a separate co-simulation setup and had to calculate all voltage magnitudes over the span of the entire data set of 365 days. 
Since the only difference between the setups is the choice of the surrogate model, all substantial changes in calculation time could be traced back to them.
To mitigate the effect of process scheduling during the experiment, the calculation was repeated numerous times.
In order to check the generalization of the benchmark's results to a different simulation model, we experimented with both the CIGRE LV and LV-rural3 simulation models. 
For the CIGRE LV, $n=10$ repetitions were performed. 
Since the calculation time of the LV-rural3 was substantially longer than that of the CIGRE LV, only $n=3$ repetitions were performed.
For every single one of those $n$ independent simulation-runs the calculation time was logged, so that we could compute the mean calculation time and its variance. 
% par 

We conducted an analysis of variance (ANOVA) to test whether the differences between the results of the models were significant.
To verify the assumptions for the ANOVA, we started to test for homogeneity of variances with Levene's test, which was not significant for neither of  grids, i.\,e., with ($F(6, 63) = 2.12, p \geq 0.05$) for CIGRE LV and ($F(6,14) = 2.09, p \geq 0.05 $) for LV-rural3, homogeneity of variances was given.
Next, we tested for normal distribution with the Shapiro-Wilk test, but this test's results were significant -- ($W=.7,p<.001$) for CIGRE LV and ($W=.66, p < .001$) for LV-rural3 -- and, thus, normality was not given.
However, since the sample size $n$ was equal for all models, we relied on the robustness of ANOVA against violations of the assumption of normal distribution with equal sample sizes \cite[p.~512]{ramachandran2014mathematical}.
After that, we conducted a two-sided independent samples Welch's t-test on every possible pair of surrogate models. 
With the results of these tests it was possible to determine whether the average calculation time differs between two surrogate models.
To reject the null hypothesis that the calculation times are identical, the tests are conducted towards a significance level of $\alpha = .05$. 
However, since we had more than two independent groups ($k = 7$), we applied the Bonferroni correction to prevent accumulation of alpha error.
% par

Additionally, the measured calculation times were used to calculate a SUF in relation to the simulation model. 
The SUF measures how much lower the surrogate models' calculation time $t_{sur}$ is compared to the simulation models' calculation time $t_{sim}$ over all $n$ simulation runs.

\begin{equation}\label{eq:suf}
\mathit{SUF} = \frac{\frac{1}{n} \sum\nolimits_{i=1}^{n} t_{sim, i}}
{\frac{1}{n} \sum\nolimits_{j=1}^{n} t_{sur, j}}
\end{equation}

As it was the case with the first experiment, we repeated the second one on the LV-rural3 grid model. 
Since it was approximately three times as big as the CIGRE-LV grid model, the results could show whether the results generalize to grid models of different sizes. 
Due to the high number of variables involved, the comparison of the results was conducted in a qualitative manner.

\subsection{Results}
\label{sec:results}

\subsubsection{Results of Experiment 1}
\label{sec:resultsexp1}

\begin{figure}[t!]
	\centering
	\includegraphics[width=0.9\textwidth]{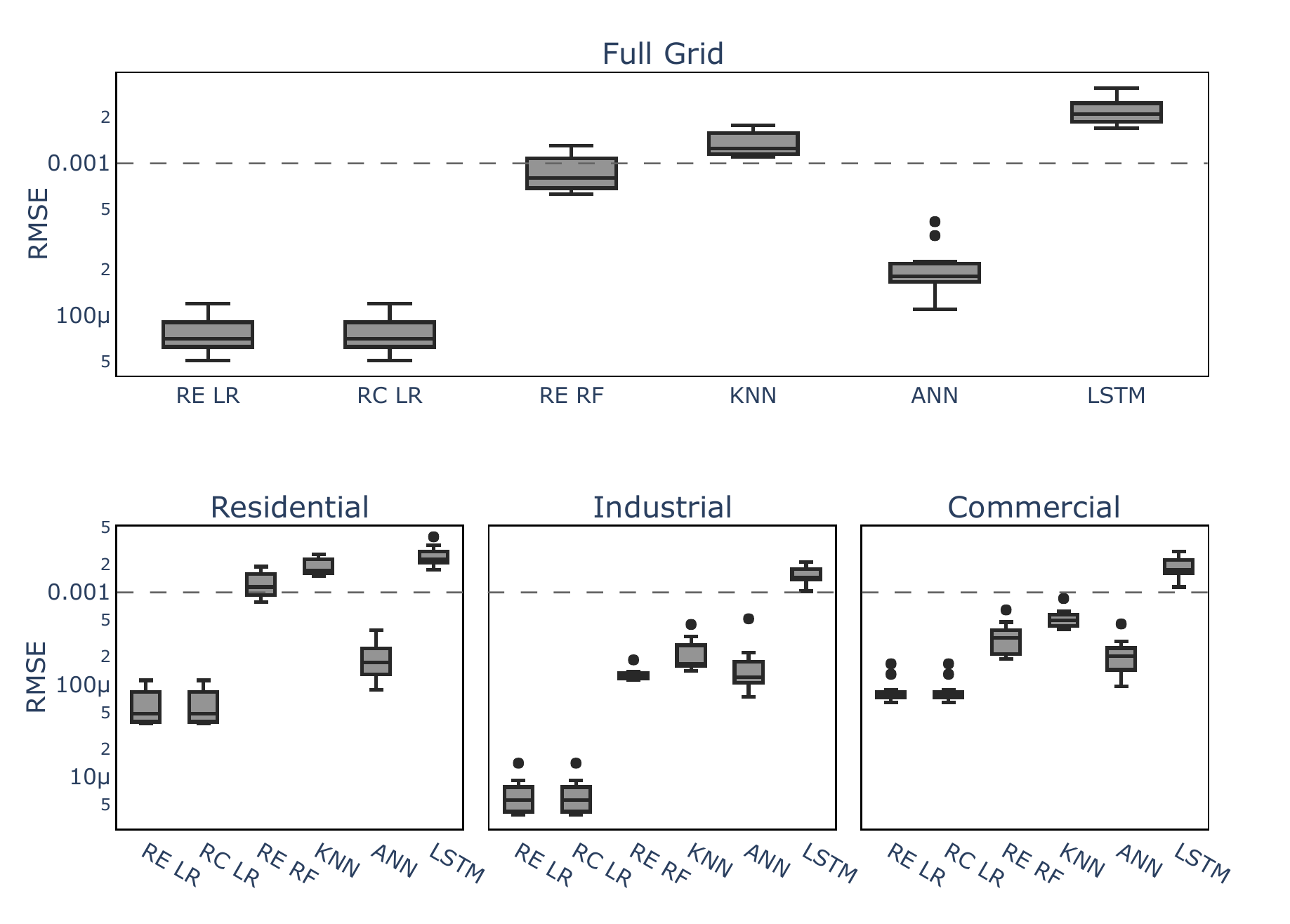}
	\caption{Boxplot of the monthly RMSE values for the evaluated surrogate models on the CIGRE LV. The dots mark outliers.	The full grid values show the average RMSE over all buses, while the lower charts consider only their respective subgrid. Due to large relative differences in RMSE values, a logarithmic scale was chosen.
	}
	\label{fig:cigrermse}
\end{figure}

The results of the first experiment on the CIGRE LV, see \autoref{fig:cigrermse}, clearly showed large differences in the obtained error values between the different surrogate models.
Especially the two models based on LR and the ANN achieved error values well below the defined criterion of an RMSE lower than 0.001, the LR even reaching RMSE values as low as 0.0001. 
Since the voltage magnitudes are given in the $pu$-system, this corresponds to error margins of \SI{0.1}{\percent} and \SI{0.01}{\percent}, respectively. 
The RF model reached an RMSE value just below the threshold of 0.001, while the k-NN model reached a value just above it. 
The LSTM, on the other hand, was not able to achieve a sufficiently low error value on the full grid. 

A closer evaluation of the RMSE values showed that the surrogate models' performance varied strongly between the different subgrids of the grid model. 
In the commercial and industrial subgrids, most models achieved lower error rates than in the residential subgrid. 
Due to the weaker results in the residential subgrid, the only surrogate models that reach satisfactory results in every subgrid were the surrogate models based on LR and the ANN. 
However, both of them stayed well below the defined threshold.
% par

\begin{figure}[t!]
	\centering
	\includegraphics[width=0.9\textwidth]{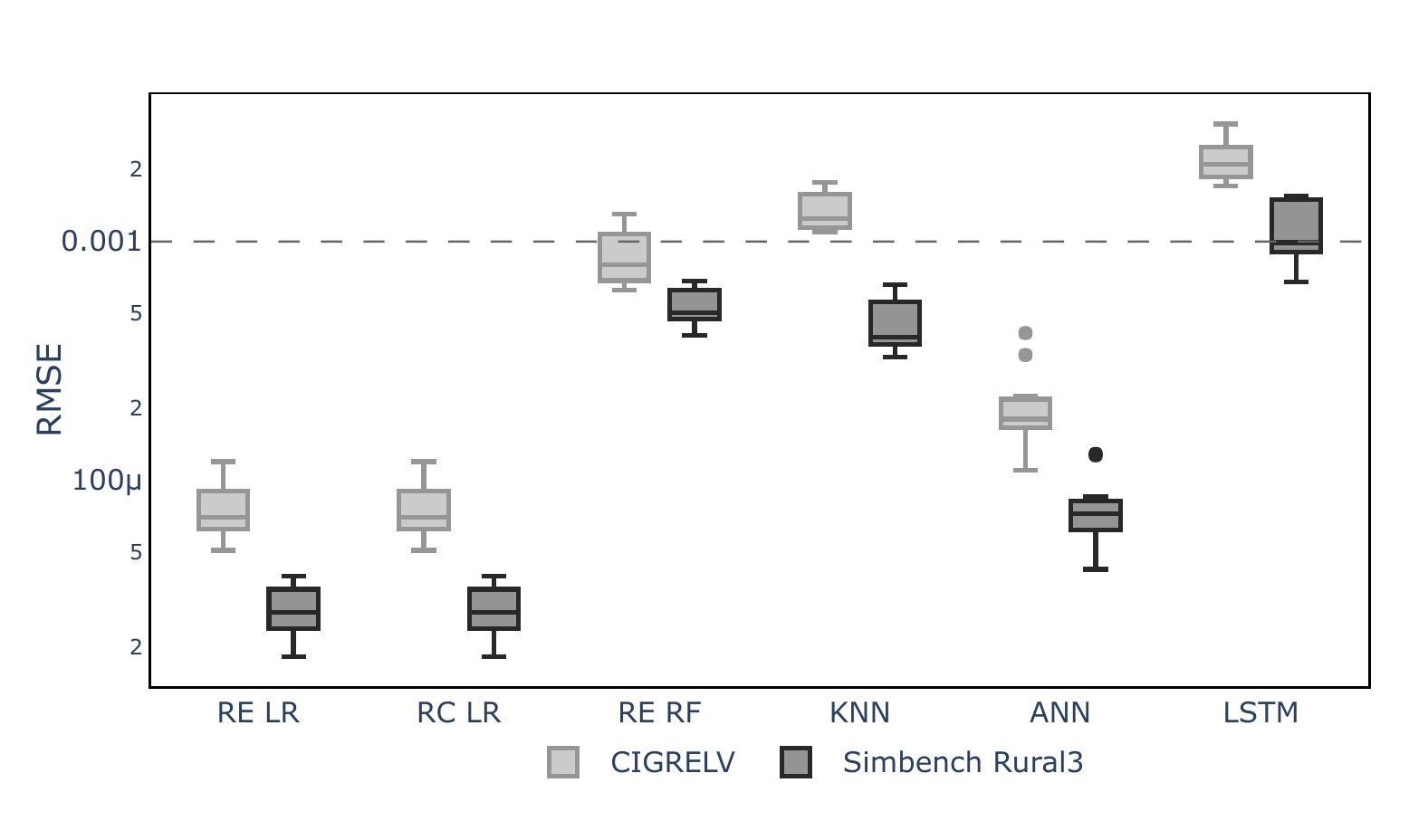}
	\caption{Monthly RMSE values for the evaluated surrogate models on the LV-rural3 grid model. The results of the CIGRE LV are displayed in lighter gray for comparison. The dots mark outliers.
	}
	\label{fig:rural3rmse}
\end{figure}

When conducted on the second grid model -- the LV-rural3, see the results in \autoref{fig:rural3rmse} -- the surrogate models generally reached a lower RMSE value than they did on the CIGRE LV. 
Despite this apparent change in the results, the general order of surrogate models remained largely the same. 
The models based on LR still reached the best results, followed once again by the ANN. 
The RF and k-NN models reached fully satisfactory results this time, while the LSTM model still struggled to stay below the threshold.

\begin{figure}[t!]
	\centering
	\includegraphics[width=0.9\textwidth]{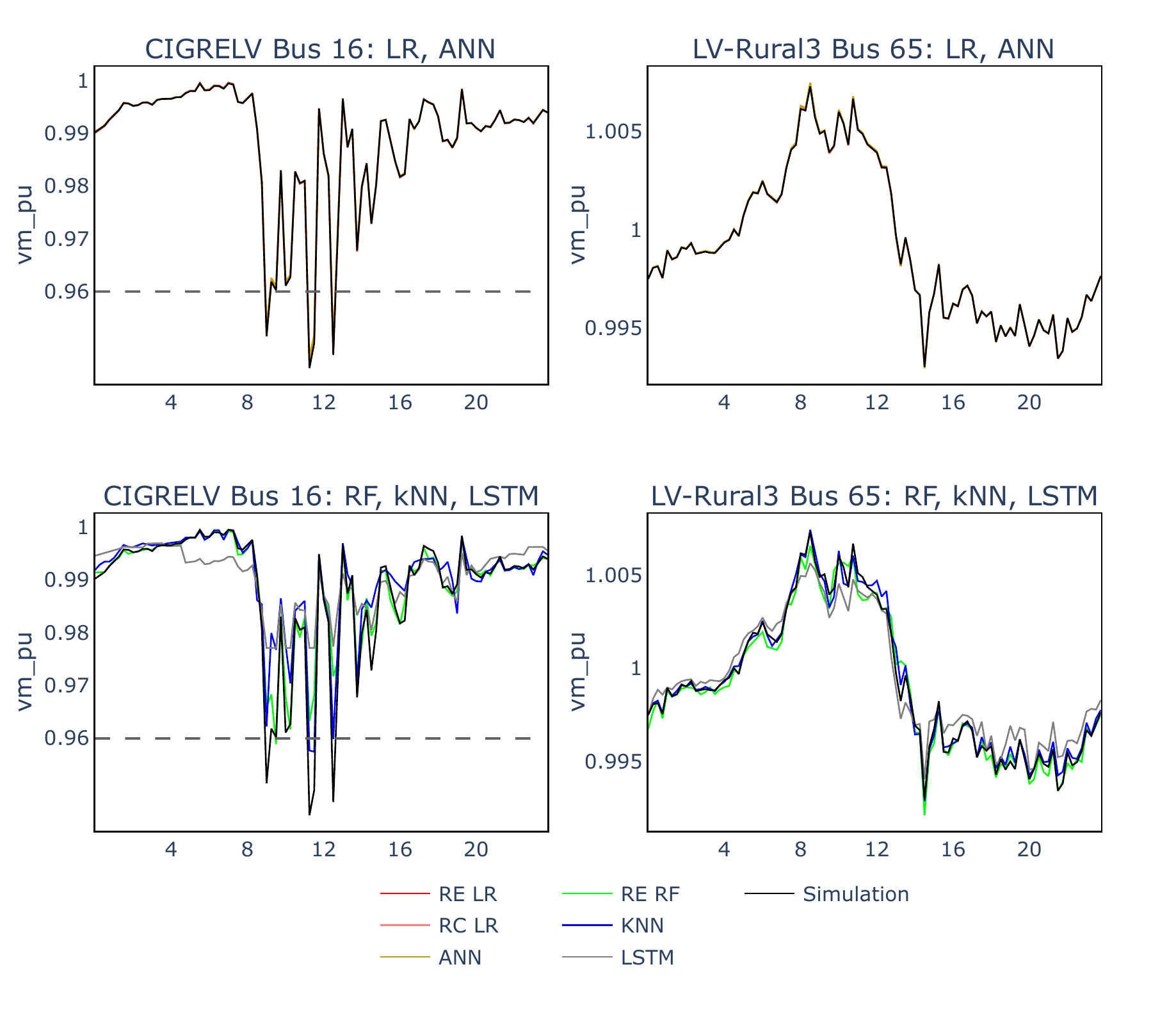}
	\caption{Comparison of simulated vm\_pus and surrogate models' predictions. The left column shows a results for one day in the residential subgrid of the CIGRE LV, while the right column shows the same for a bus in the LV-rural3.  Both buses obtained the highest error values in their respective grid models.
	}
	\label{fig:exp1comp}
\end{figure}

\autoref{fig:exp1comp} shows the differences between the surrogate models' results. 
Especially in the rapidly changing \glspl*{vmpu} in the CIGRE LV, the less accurate models struggled to accurately map the simulation models' behavior. 
The models based on LR and the ANN, on the other hand, barely showed any visible discrepancies.
% par

Since overall results were better than they were on the smaller CIGRE LV, no negative impact on the prediction results by changes of the grid topology was apparent. 
Therefore, the results indicated that the use of surrogate models is robust towards changes in grid size.

We conclude the same statement towards the inclusion of a changing phase angle. 
While $cos(\varphi)$ was fixed at 0.9 in the CIGRELV, it was not constrained in the LV-rural3.
However, this did not affect the surrogate models' ability to accurately map the simulation model's behavior.
Additionally, the inclusion of volatile distributed energy generation in the form of PV generators did not have a negative effect on the surrogate models' accuracy. 
Feeding the generated energy into the surrogate model in the same way as the loads appeared to be a viable solution to the inclusion of energy generated in a decentralized manner. 
% par

All things considered, the experiment showed that results concerning accuracy obtained from CIGRE LV generalized remarkably well to grids with different factors. 
We could also show that there are large differences in the surrogate models' abilities to map the underlying grid simulation model. 
Especially the parametric models based on LR and the ANN performed well. 
The LSTM -- despite being an extension of the neural network well suited to time series -- did not manage to reach acceptable results on either grid model. 
A possible explanation would be the large 15-minute intervals between measurements. 
Due to the more rapid fluctuations in the voltage magnitudes and loads, the greater flexibility of the LSTM seemed to be more of an obstacle than a help due to the limited amount of relevant information in the preceding time steps.

\subsubsection{Results of Experiment 2}
\label{sec:resultsexp2}

The results obtained from the second experiment are meant to showcase the differences of calculation time between the surrogate models and the simulation model.
In the ANOVA, the average calculation times over ten repetitions showed significant differences between the models ($F(6, 63) = 11291.78, p < .001$).
Therefore, we conducted Welch's t-test, which is robust to a violation of normality \cite{hansen2005using}, to determine pairwise differences.
The results -- see \autoref{tab:exp2tcigre} -- show significant differences between almost all model combinations. 
The only exceptions were marked by the pairs (Sim, LSTM) and (RC LR, ANN), which had no significant differences in the test. 
As with the first experiment, the ANOVA for LV-rural3 showed significant differences between the models concerning calculation time and we could conduct Welch's $t$-test. 
The results in \autoref{tab:exp2tcigre} showed significant differences between most of the model combinations, except for the pairs (Sim, RE RF), (RE LR, RC LR), (RC KR, k-NN), and (ANN, LSTM).

\begin{table}[t!]
	\centering
	\caption{Results from the independent samples Welch's t-test for both grids. Shown are the t-statistics and the corresponding degree of freedom ($df$). Reported p-values were Bonferroni-corrected.
	}
	\begin{tabular}{llcc|cc}
		\toprule
		& & \multicolumn{2}{c}{CIGRE LV} &  \multicolumn{2}{c}{LV-rural3} \\
		Model A & Model B & $df$ & $t$ &  $df$ & $t$ \\ 
		\midrule
		Sim & RE LR & 10.4 & 56.81*** & 2.0 & 25.54* \\
		Sim & RE RF & 16.7 & -78.38*** & 2.1 & -13.03 \\
		Sim & RC LR & 11.2 & 60.83*** & 2.0 & 26.32* \\
		Sim & k-NN & 11.2 & 63.75*** & 2.0 & 26.02* \\
		Sim & ANN & 9.7 & 64.04*** & 2.0 & 27.77* \\
		Sim & LSTM & 10.4 & 1.38 & 2.0 & 27.58* \\
		RE LR & RE RF & 9.8 & -139.03*** & 2.3 & -302.50*** \\
		RE LR & RC LR & 17.3 & 11.94*** & 3.8 & 17.25** \\
		RE LR & k-NN & 17.3 & 18.78*** & 2.2 & 13.02 \\
		RE LR & ANN & 16.3 & 17.97*** & 3.5 & 52.94*** \\
		RE LR & LSTM & 18.0 & -143.72*** & 2.6 & 53.87** \\
		RE RF & RC LR & 10.2 & 141.40*** & 2.2 & 312.91*** \\
		RE RF & k-NN & 10.2 & 143.63*** & 2.0 & 317.89*** \\
		RE RF & ANN & 9.4 & 145.17*** & 2.1 & 326.95*** \\
		RE RF & LSTM & 9.8 & 96.85*** & 2.0 & 329.09*** \\
		RC LR & k-NN & 18.0 & 6.23*** & 2.2 & -10.71 \\
		RC LR & ANN & 14.5 & 2.22 & 3.9 & 40.01*** \\
		RC LR & LSTM & 17.2 & -140.06*** & 2.9 & 40.33*** \\
		k-NN & ANN & 14.5 & -5.40** & 2.3 & 72.84** \\
		k-NN & LSTM & 17.2 & -146.90*** & 3.0 & 102.65*** \\
		ANN & LSTM & 16.4 & -183.52*** & 3.2 & -7.44 \\
		\bottomrule
	\end{tabular}
	\\
	*** $p < .001$, ** $p < .01$, * $p < .05$
	\label{tab:exp2tcigre}
\end{table}

\autoref{tab:calctimes} illustrates that almost all models resulted in a considerable reduction of calculation time in comparison to the simulation model.
The LSTM had almost the same calculation time as the simulation model, the RF model was unable to decrease the calculation time and was, in fact, slower than the simulation model. 
The fastest one of those was the k-NN model, which reached an average SUF of 3.74. 
It was closely followed by the ANN (3.43) and the LR models (3.32 and 2.78).
For comparison, the reference model reached SUFs between 2.5 and 2.7 in a very similar experiment on the same grid model \cite[p.~14]{balduin2019towards}.

\begin{table}[t!]
	\centering
	\caption{Overview of RMSE, mean calculation time (Calc. [s]), its standard deviation (StD.), and the models' speed up factor (SUF) relative to the simulation model (Sim).
	}
	\begin{tabular}{lcccc|cccc}
		\toprule
		& \multicolumn{4}{c}{CIGRELV} & \multicolumn{4}{c}{LV-rural3} \\
		{}         & RMSE     &Calc. [s]&   StD. & SUF	  & RMSE     &Calc. [s] &   StD. & SUF \\ 
		\midrule
		Sim & 	-     &1354.45 	&  46.45 &  1.00  &     -    &  6388.01 & 375.63 &  1.00 \\
		RE LR      & \textbf{7.57e-05} & 486.64  &  13.21 &  2.78  & \textbf{2.91e-05} &   846.36 &  13.19 &  7.55 \\
		RC LR      & \textbf{7.57e-05} & 407.48  &  16.35 &  3.32  & \textbf{2.91e-05} &   678.38 &  10.50 &  9.42 \\
		RE RF      & 8.80e-04 &3275.52  &  62.04 &  0.41  & 5.36e-04 &  9234.81 &  46.19 &  0.69 \\
		k-NN        & 1.34e-03 & \textbf{362.26}  &  16.26 &  \textbf{3.74}  & 4.55e-04 &   745.28 &   2.58 &  8.57 \\
		ANN        & 2.08e-04 & 394.39  &   9.45 &  3.43  & 7.75e-05 &   \textbf{363.11} &   8.71 & \textbf{17.59} \\
		LSTM       & 2.19e-03 &1333.35  &  13.13 &  1.02  & 1.13e-03 &   406.46 &   5.10 & 15.72 \\
		\bottomrule
	\end{tabular}
	\label{tab:calctimes}
\end{table}

\begin{figure}[t!]
	\centering
	\includegraphics[width=0.9\textwidth]{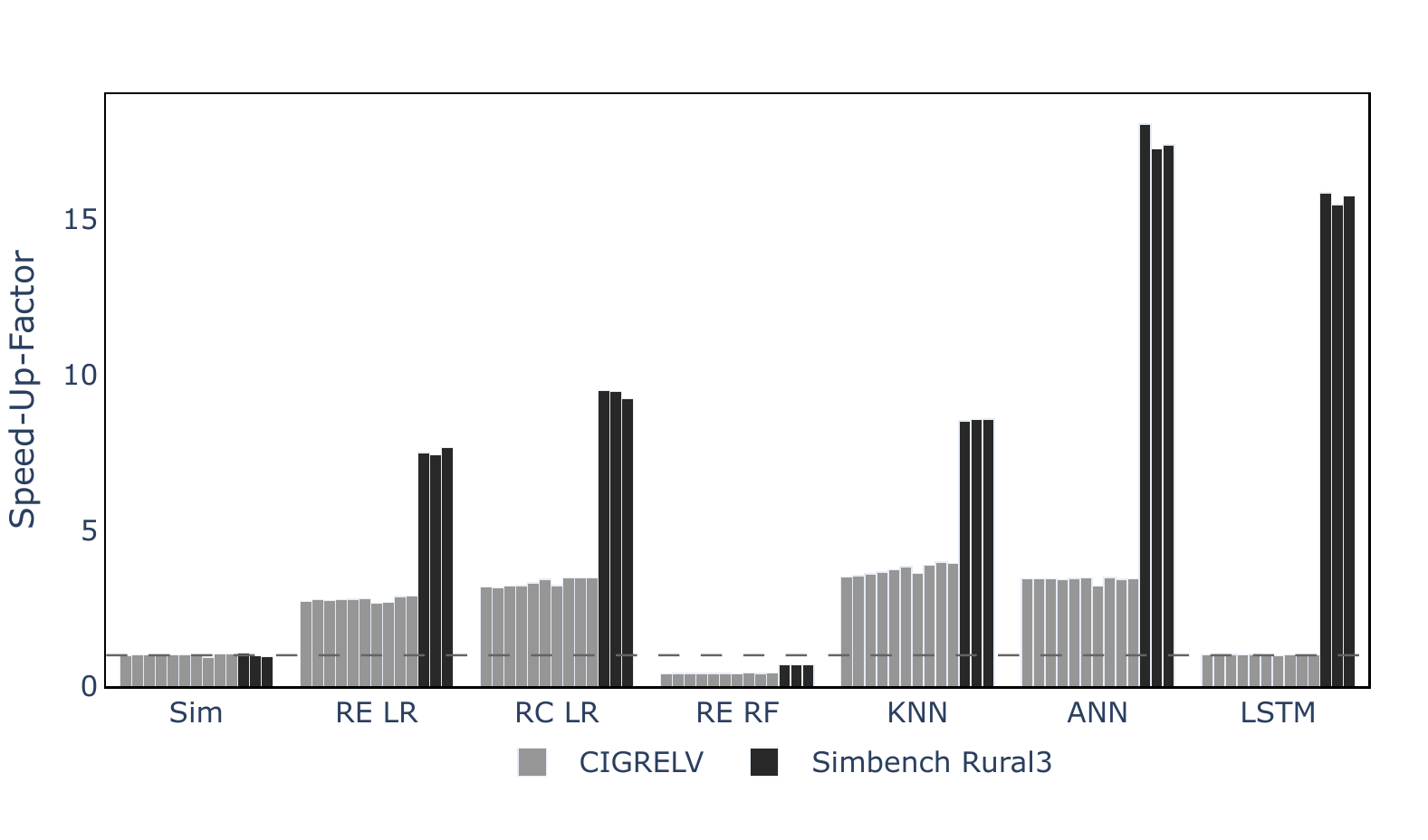}
	\caption{Comparison of speed- up factors over all calculation times. The bars on the far left are the simulation models (Sim) and the reference for the other models. Higher is better.}
	\label{fig:speedcomp}
\end{figure}

Since the LV-rural3 model was far larger than the CIGRE LV, the simulation model showed a far longer calculation time, illustrated in \autoref{tab:calctimes}. 
The same holds true for the majority of the surrogate models, which -- except for the ANN and the LSTM -- also took longer to calculate the relevant outputs. 
However, the increase in calculation time was less pronounced for the surrogate models. 
Therefore, all surrogate models reached higher SUFs when applied to the LV-rural3 than when applied to the smaller CIGRE LV.
This effect was especially strong for the ANN and the LSTM, allowing them to reach SUFs of 17.59 and 15.72, respectively. 
With SUFs between 7.5 and 9.5, the remaining surrogate models achieved satisfactory results, as well. 
The results from both grid models, illustrated in \autoref{fig:speedcomp}, indicate that some models reacted more strongly to a larger grid model than others. 
Especially the ANN barely reacted to the larger size.

\section{Conclusion}
\label{sec:conclusion}

In this work, we evaluated different algorithms regarding their adequacy to be used as a surrogate model for a LV grid.
These algorithms can be categorized into ensembles of single-target models (LR and RF) and multi-target models (k-NN, ANN, and LSTM), models with low (k-NN and RF) to high (ANN and LSTM) number of hyperparameters, distance-based models (k-NN), or neural network models (ANN and LSTM).
All surrogate models were evaluated in two experiments, the first one regarding the accuracy and the second one regarding the calculation time compared to the simulation model.
Additionally, we changed some of the LV grid's parameters, namely the topology itself, the use of distributed energy generation, and the change from a constant to a varying phase angle.
The changes were made to provide an estimation on the robustness regarding different grid topologies. 
Therefore, we conducted the two experiments on two different grid models: the CIGRE LV benchmark grid and the simbench LV-rural3 grid.
% par

In the results of the first experiment we could verify that the LR-based models (RE LR and RC LR) and the ANN had a prediction error far below \SI{0.1}{\percent} on both grid models, while RF on both grids and k-NN at least on the LV-rural3 grid still achieved satisfactory results.
Only the LSTM did not pass the prediction error cutoff. 
On the CIGRE LV grid, the prediction error of the models differed significantly depending on the subgrid.
We attribute this to more regular behavior of the commercial data sets.
Furthermore, the change of the topology and other parameters in the second grid model had nearly no effect on the general order of the surrogate models.
From this, we deduced that the surrogate model algorithms were robust against parameter changes of the replaced simulation models.
% par

The results of the second experiment showed the calculation time benefit of using surrogate models.
Each surrogate model was significantly faster and provided a speed-up against the simulation model with one exception: the RF algorithm applied on the LV-rural3 was even slower than the simulation model.
This could be caused by a number of trees that was chosen too high. 
For the other models holds true that a change to a larger grid had actually increased the speed-up compared to the simulation model.
This applied in particular, to the ANN-based models\footnote{The higher calculation time of the LSTM network on the CIGRE LV can likely be attributed to the use of a second (calculation intensive) recurrent layer that was a result from hyperparameter tuning.} whose calculation time was hardly influenced by the grid size.
We conclude that a LR-based model or a k-NN model is good enough as a surrogate model for smaller grids, while ANN-based models further extended their advantage on larger grids.
Considering the results obtained from both grid models, the experiment showed that a wide variety of different surrogate models can be used to decrease the calculation time of the simulation models. 

From both experiments, we concluded that the use of an ANN as the surrogate model in the reference work \cite{balduin2019towards} was not a bad choice although the evidence was not provided in that very work.
Additionally, we improved the model architecture to provide an even lower error and a higher speed-up.
Furthermore, we could verify that a change of the grid topology and other parameters like the integration of distributed energy resources and a varying phase angle had no negative impact on the quality of the model.

\section{Outlook}
\label{sec:outlook}

We addressed some of the open questions of the reference work, but some of them still are unanswered and new questions emerged.
The tuning of hyperparameters could be further improved, e.\,g., penalizing the calculation time in the loss function, which could create a tendency towards smaller models.
This would address the question of how much effort has to be made in order to obtain a useful surrogate model and how this effort relates to the benefits.
In the reference work, there was an experiment regarding the detection of critical voltage violations, which could not be transferred to LV-rural3 since there were no voltage violations inside the data.
The problem could be circumvented by adding more data to the grid and then conducting the corresponding experiments again.
Further improvements of the model could be considered, such as the inclusion of line loadings or other outputs of the grid simulation model.
Another aspect would be to build a larger setup with a MV grid and several LV grids and investigate if the findings of this work still hold true.
% par

Finally, the issue of limited training data should be addressed and we think that two approaches are conceivable.
First, the available training data could be extended artificially by, e.\,g., the use of input distributions or bootstrapping.
Secondly, the amount of training data required could be reduced by taking into account correlations of loads and generation in the surrogate modeling process.
In our future work, we aim to address these question and, at the same time, include feedback from actual applications of our methodology in other related projects.

\section*{Acknowledgements}
Parts of this work have been developed in the context of a master thesis at the Carl v. Ossietzky University of Oldenburg and the Helmut-Schmidt University of Hamburg. We thank our supervisors Prof. Dr.-Ing Alexander Fay, Prof. Dr. Sebastian Lehnhoff and Jan-Philip Beck. 
We also thank our colleagues Johannes Gerster, Lasse Hammer, Daniel Lange and Eric Veith for their helpful comments and contributions during the preparation of this manuscript.

\bibliographystyle{unsrt}
%\bibliography{references}  

\end{document}